\documentclass[10pt,twocolumn]{article}

\usepackage[utf8]{inputenc}
\usepackage[T1]{fontenc}
\usepackage{lmodern}
\usepackage{amsmath,amssymb}
\usepackage{graphicx}
\usepackage{authblk}
\usepackage{geometry}
\usepackage{hyperref}
\usepackage{titlesec}
\usepackage{tikz}
\usepackage{caption}
\usepackage{subcaption}
\usepackage{mathrsfs}
\usetikzlibrary{positioning}
\usetikzlibrary{positioning,calc} 
\usetikzlibrary{shapes.geometric}
\usepackage[numbers,sort&compress]{natbib}
\usepackage{appendix}
\hypersetup{
	colorlinks=false,     
	linkbordercolor={0.7 0.85 1}, 
	citebordercolor={0 0.4 0},
	pdfborderstyle={/S/S/W 1}    
}

\titleformat{\section}
{\normalfont\bfseries\large}{\thesection.}{0.5em}{}

\titleformat{\subsection}
{\normalfont\bfseries}{\thesubsection.}{0.5em}{}

\makeatletter
\renewcommand\maketitle{
	\begin{center}
		{\LARGE\bfseries \@title \par}
		\vskip 0.7em
		{\large \@author \par}
		\vskip 0.7em
		{\@date}
	\end{center}
	\vspace{1.3em}
}
\makeatother

\title{Causal Structure for Generalized Spinfoams}

\author[1]{Carlos E. Beltrán.\thanks{carlosbeltran27111992@gmail.com}}

\affil[1]{
	Centro de Ciencias Matemáticas, 
	Universidad Nacional Autónoma de México, 
	Morelia, Michoacán C.P. 58089, México. \\
	Instituto de Física, 
	Universidad Nacional Autónoma de México, 
	Ciudad de México C.P. 04510, México. 
}

\date{April 21, 2026}
\begin{document}
	
	\twocolumn[%
	\maketitle
	
	\begin{center}
		{\bfseries\large\scshape Abstract}
	\end{center}
	
	\begin{center}
		\begin{minipage}{0.75\textwidth}
			We investigate the role of causality in the generalized EPRL spinfoam model introduced by Kaminski, Kisielowski, and Lewandowski (EPRL-KKL). We first propose a definition of causal structure on an arbitrary 2-complex and analyze how causal orientations can be assigned to its 1-skeleton and 2-skeleton. We identify a criterion that determines when an orientation of the 2-skeleton induces a consistent causal structure on the 1-skeleton. This characterization naturally involves tools from graph theory and linear algebra over the Galois field $\mathbb{F}_2$. Using these results, we introduce a causal vertex amplitude that generalizes previous proposals by Bianchi--Dussaud and Bianchi--Chen--Gamonal. We study its asymptotic behavior and revisit the interpretation of the two critical points appearing in the semiclassical analysis of the EPRL and EPRL-KKL spinfoam models. Finally, we discuss several directions in which the causal amplitude introduced here may provide new insights for future developments, like the cosine problem and the presence of irregular light cone structures. 
		\end{minipage}
	\end{center}
	
	\vspace{1.0em}
	]  
	
	\section{Introduction}
	
	The spinfoam framework is a path-integral approach to quantum gravity within the program of loop quantum gravity \cite{Perez:2012wv, Livine:2010zx, R-V:2014, Ashtekar:2021}. The most successful candidate to date is the EPRL model \cite{Engle:2007wy, Rovelli:2011eq}. This proposal was later generalized to include graphs with arbitrary valence through the so-called EPRL-KKL model \cite{Kaminski:2010, Kaminski:2009cc}. Subsequently, this model was derived as a quantization of constrained BF theory discretized over an arbitrary 2-complex \cite{Ding:2010fw}. Its applications include the calculation of black-to-white hole tunneling transition amplitudes, spinfoam cosmology, and spinfoam renormalization \cite{Christodoulou:2024, Bianchi:2018rem, DAmbrosio:2021, Soltani:2021, Bianchi:2010cosmo, Bahr:2016a, Bahr:2016b, Bahr:2018}.
	
	The role of causality in the spinfoam framework has been studied since the early literature \cite{Markopoulou:1997, Markopoulou:1998, Markopoulou:2000, Gupta:2000, Livine:2003, Oriti:2005, Pfeiffer:2003, Hawkins:2003, Oriti:2006b, Livine:2007, Cortes:2016, Immirzi:2016, Jercher:2022}. Very recently, causality in the EPRL model has been analyzed and new proposals for causal vertex amplitudes have been introduced \cite{Bianchi:2024, Bianchi:2026rjd}. In this article we extend these studies to the EPRL-KKL model. In particular, we uncover novel connections between causality and algebraic graph theory. Following the lines developed in \cite{Bianchi:2024} and \cite{Bianchi:2026rjd}, we propose new causal vertex amplitudes in the context of an arbitrary 2-complex. We analyze the semiclassical limit of these amplitudes and show that they select a single critical point, in accordance with the structure identified in \cite{Dona:2020}. Finally, we discuss the interpretation of these causal amplitudes and their role in clarifying several aspects of the spinfoam approach.

	\section{Discrete Causal Structure}
	\subsection{Discretization}
	An oriented 2-complex is defined as an ordered quartet $\mathcal{C}:=(V,E,F,\partial)$ where:
	\begin{itemize}
		\item $V$, $E$ and $F$ are sets whose elements are called vertices $v\in V$, edges $e\in E$ and faces $f\in F$. 
		\item $\partial$ is a relation that associates an ordered pair of vertices $(s(e),t(e))$ (called the \textit{source} and \textit{target}) to each edge $e$ and a finite sequence of edges $\left\{e_{k}^{\epsilon_{e_{k}f}}\right\}$ to each face $f$, in such a way that $t(e_k)=s(e_{k+1})$, $t(e_n)=s(e_1)$ and $\epsilon_{ef}=\pm1$. The quantity $\epsilon_{ef}$ is called the orientation of the edge $e$ with respect to the face $f$.
	\end{itemize}
	Here $\partial f$ denotes the set of oriented edges, or equivalently the ordered set of vertices around a face $f$. Similarly, $\partial e$ denotes the set of vertices that bound the edge $e$. We write $\partial v$ to denote the set of edges incident at $v$, or the set of faces $f$ such that $v\in\partial f$. The distinction will be clear from the context.\\
	\indent The boundary graph of $\mathcal{C}$ will be denoted by $\gamma$. The boundary graph of a single vertex will be denoted by $\gamma_{v}$.
	
	\indent For each vertex, $N_{v}^{e}$ and $N_{v}^{f}$ denote the number of edges $e\in \partial v$ and faces $f\in\partial v$, respectively. The quantity $N_{v}^{e}$ will be called the valence of the vertex $v$. A pair $(e,v)$ with $v\in\partial e$ is called an admissible pair, and $\mathcal{E}\times\mathcal{V}$ will denote the set of admissible pairs.
	
	\subsection{Discrete causality}
	
	\subsubsection{Causality in Minkowski spacetime}
	
	Let $\mathcal{M}$ denote the four-dimensional Minkowski spacetime with metric $g=\text{diag}(+\eta,-\eta,-\eta,-\eta)$, where $\eta=\pm 1$. Notice that we are working with an arbitrary Minkowski signature, as in \cite{Bianchi:2024}. Given a vector $u\in\mathcal{M}$, we can classify it into three different types:
	\begin{equation}
		\text{sgn}\left(g(u,u)\right)=
		\begin{cases}
			\eta & \text{if } u \text{ is timelike} \\
			0    & \text{if } u \text{ is lightlike} \\
			-\eta & \text{if } u \text{ is spacelike}.
		\end{cases}
	\end{equation}
	Each set will be called a \textit{causal class}.
	
	Let $\tau$ be the set of timelike vectors in $\mathcal{M}$. It is possible to define an equivalence relation on $\tau$ as follows: if $u,w\in\tau$, then $u\sim w$ if and only if $\text{sgn}\left(g(u,w)\right)=\text{sgn}\left(g(w,w)\right)$ \cite{Naber:2012}. This relation partitions $\tau$ into exactly two equivalence classes that we denote by $\tau^{+}$ and $\tau^{-}$. Choosing a reference timelike vector $u_{0}\in\mathcal{M}$, we say that an element $w\in\tau$ belongs to $\tau^{+}$ if
	\begin{equation}
		\text{sgn}\left(g(u_{0},w)\right)=\eta.
	\end{equation}
	We will call $\tau^{+}$ and $\tau^{-}$ the future-directed and past-directed timelike vectors, respectively.
	\subsubsection{Causal structure on $\mathcal{C}$}
		
	In a discretization dual to a triangulation, it is possible to associate two vectors $N_{(e,v_1)}$, $N_{(e,v_2)}\in\mathcal{M}$ to each edge $e\in E$, with $v_1, v_2\in\partial e$. These elements correspond to the outward-pointing unit normal vectors of the tetrahedron dual to $e$, as seen from the 4-simplices dual to $v_1$ and $v_2$, respectively. Moreover, $N_{(e,v_1)}$ and $N_{(e,v_2)}$ belong to the same causal class for all $e$. This allows to associate a causal structure to the 2-complex $\mathcal{C}$ in this particular case \cite{Bianchi:2024}.
	
	\indent In a more general setting, we will define a discrete causal structure on an arbitrary 2-complex  $\mathcal{C}$ as a function $\mathcal{X}:\mathcal{E}\times\mathcal{V}\rightarrow \mathcal{M}$ such that:
	\begin{itemize}
		\item If $v_{1}, v_{2}\in\partial e$, then $\mathcal{X}_{(e,v_1)}$ and $\mathcal{X}_{(e,v_2)}$ belong to the same causal class.
	\end{itemize}
	Two discrete causal structures $\left\{\mathcal{X}_{(e,v)}\right\}$ and $\left\{\mathcal{Y}_{(e,v)}\right\}$ on $\mathcal{C}$ will be said to be equivalent if and only if, for every admissible pair $(e,v)$, there exists an orthochronous proper Poincaré transformation $\Lambda_{(e,v)}$ such that
	\begin{equation}
		\mathcal{X}_{(e,v)}=\Lambda_{(e,v)} \rhd \mathcal{Y}_{(e,v)}.
	\end{equation}
	
	In a discretization dual to a triangulation, it is possible to choose a local time orientation at each vertex $v$ by classifying the spacelike tetrahedra as past- or future-directed \cite{Bianchi:2024}. We can generalize this notion as follows.
	
	Let $\tau_{v}$ be the set of timelike vectors associated with the edges incident at a vertex $v$. As in \cite{Bianchi:2024}, we can associate a temporal orientation to each vertex. Given a vertex $v\in V$, we divide $\tau_{v}$ into two equivalence classes $\tau_{v}^{+}$ and $\tau_{v}^{-}$ by the relation
	\begin{equation}\label{equivalence}
		\begin{aligned}
			\mathcal{X}_{(e_1,v)} \sim \mathcal{X}_{(e_2,v)} \iff & \text{sgn}\left(g(\mathcal{X}_{(e_1,v)}, \mathcal{X}_{(e_2,v)})\right) \\
			& = \text{sgn}\left(g(\mathcal{X}_{(e_1,v)}, \mathcal{X}_{(e_1,v)})\right).
		\end{aligned}
	\end{equation}
	
	We can choose a local time orientation at the vertex $v$ by specifying which class is called past and which one is called future. This construction generalizes the notion of time orientation at a vertex introduced in \cite{Bianchi:2024}.
	
	Given an edge $e\in E$, we say that $e$ is timelike, spacelike, or lightlike if $\mathcal{X}_{(e,v)}$ is respectively timelike, spacelike, or lightlike for any $v\in\partial e$.
	
	We say that $\mathcal{C}$ is time-orientable if we can consistently choose a time orientation at each vertex in such a way that every edge $e$ has opposite time directions with respect to $s(e)$ and $t(e)$. If such orientation has been made, we say that $\mathcal{C}$ is time-oriented.
	
	Given two edges $e_{1}, e_{2}$ incident at a vertex $v$, if $e_{1}$ and $e_{2}$ belong to the same class $\tau_{v}^{+}$ or $\tau_{v}^{-}$, they are said to be cochronal. Otherwise, they are said to be antichronal.
	
	In the remainder of this article we assume that all edges $e$ are timelike. The time orientation of each edge $e$ with respect to a vertex $v$ will be encoded in a variable $\epsilon_{v}(e)$ defined as
	\begin{equation}
		\epsilon_{v}(e):=
		\begin{cases}
			-1 & \text{if } e \text{ is past-directed with respect to } v, \\
			+1 & \text{if } e \text{ is future-directed with respect to } v.
		\end{cases}
	\end{equation}
	
	If $v_{1}, v_{2}\in\partial e$ and $\mathcal{C}$ is time-oriented, then $\epsilon_{v_1}(e)=-\epsilon_{v_2}(e)$. 
	
	In the rest of this article we will always assume that $\mathcal{C}$ is time-oriented.
	
	With this structure, the 1-skeleton acquires the structure of a causal set, in the same way as in \cite{Bianchi:2024, Bianchi:2026rjd}. However, the resulting ordering on the set of vertices does not satisfy additional conditions, such as acyclicity, that would prevent the existence of closed timelike loops \cite{Oriti:2005}.
	
	This notion of orientation should not be confused with the standard orientation of edges commonly used as auxiliary data in the definition of spinfoam amplitudes \cite{Perez:2012wv, Livine:2010zx, R-V:2014, Ding:2010fw}. In the usual construction, edge orientations are face-dependent: a given edge may have different orientations depending on the face with respect to which it is considered \cite{R-V:2014, Ding:2010fw}. By contrast, the causal orientation introduced here is an intrinsic property of each edge that is independent of any face.
	
	As discussed in the literature, causal spinfoam amplitudes are defined in such a way that they incorporate this orientation in a non-trivial way, leading to orientation-dependent transition amplitudes for quantum gravity \cite{Oriti:2005, Oriti:2006b}. In the Lorentzian setting, such amplitudes admit an interpretation as causal transition amplitudes, playing a role analogous to that of the Feynman propagator in standard quantum field theory\cite{Gupta:2000, Livine:2003, Oriti:2005, Immirzi:2016, Teitelboim:1982}.
	
	\subsubsection{Causal Structure in the 2-skeleton} 
	
	It is possible also to associate a causal orientation to the 2-skeleton. If $v\in V$ and $f\in \partial v$, the pair $w:=(v,f)$ will be called a wedge. Given a wedge, there are two edges $e_{1}, e_{2}\in\partial v$ such that $e_{1}, e_{2}\in\partial f$. We define the dihedral angle of $w$ as
	
	\begin{equation}
		\theta_{w}:=\text{sgn}\left(g(\mathcal{X}_{(e_1,v)}, \mathcal{X}_{(e_2,v)})\right)\cosh^{-1}\!\left(|g(\mathcal{X}_{(e_1,v)}, \mathcal{X}_{(e_2,v)})|\right).
	\end{equation}
	Then
	\begin{equation}
		\text{sgn}\left(\theta_{w}\right) = 
		\begin{cases}
			+\eta & \text{if } e_1, e_2 \text{ are cochronal}, \\
			-\eta & \text{if } e_1, e_2 \text{ are antichronal}.
		\end{cases}
	\end{equation}
	
	We define the orientation of the wedge $w=(v,f)$ as $\epsilon_{v}(f):=\text{sgn}\left(\theta_{w}\right)$. Then we have
	
	\begin{equation}\label{wedgeor}
		\epsilon_{v}(f)=\eta\,\epsilon_{v}(e_1)\epsilon_{v}(e_2).
	\end{equation}
	
	The orientation of a wedge $w=(v,f)$ is essentially the product of the orientations of the two edges $e_{1}, e_{2}\in\partial f$. A natural question is whether, given all the $\epsilon_{v}(f)$, it is possible to invert the system of equations defined by \eqref{wedgeor} in order to obtain all the $\epsilon_{v}(e)$. The case of a 5-valent vertex, which corresponds to a 4-simplex, was analyzed in \cite{Bianchi:2024}. Here the situation becomes more interesting, especially because the valence $N_{v}^{e}$ is arbitrary, and not every pair of edges $e_{1}, e_{2}\in\partial v$ forms a face $f$. In this article we assume that the 2-complex $\mathcal{C}$ is such that the boundary graph of each vertex $v$ is 3-link connected\footnote{A graph is 3-link connected if any bipartition of the nodes cannot be disconnected by cutting only two links. See \cite{Dona:2020}.}. The reasons for this choice will become clear in the next section.\\
	
	\indent If every vertex $v\in V$ is such that its boundary graph is 3-link connected, then $N_{v}^{e}\geq 4$ and $\gamma_{v}$ is connected and contains cycles.\\
	
	\indent Given a vertex $v$, we label the edges $e\in\partial v$ with $a=1,\dots, N_{v}^{e}$, and the oriented links of $\gamma_{v}$ with $(ab)$. In order to determine whether we can invert \eqref{wedgeor}, we introduce the \textit{causal graph of a vertex $v$}, denoted by $\mathcal{G}_{v}$. The causal graph of a vertex is constructed by associating a node to each unknown $\epsilon_{v}(e)$ and an edge to each $\eta\epsilon_{v}(f)$, connecting the nodes according to the structure of the system \eqref{wedgeor}. In the causal graph $\mathcal{G}_{v}$, a cycle is a sequence of edges and nodes that closes. If the causal structure in the 1-skeleton $\epsilon_{v}(e)$ is given, then it can be shown that the product of orientations $\epsilon_{v}(f)$ around a cycle in the causal graph satisfies
	
	\begin{equation}\label{cycle}
		\prod_{f\in\text{cycle}}\epsilon_{v}(f)=\eta^{s},
	\end{equation}
	where $s$ is the number of faces around the cycle.
	
	\begin{figure}[htbp]
		\centering
		\begin{tikzpicture}[
			scale=0.6, 
			transform shape,
			every node/.style={draw, ellipse, minimum width=1.8cm, inner sep=2pt},
			label_edge/.style={draw=none, rectangle, font=\small, inner sep=1pt}
			]
			
			\node (e1) at (90:3)  {$\epsilon_{v}(e_1)$};
			\node (e2) at (18:3)  {$\epsilon_{v}(e_2)$};
			\node (e3r) at (-54:3) {$\epsilon_{v}(e_3)$}; 
			\node (e3l) at (234:3) {$\epsilon_{v}(e_4)$}; 
			\node (e3m) at (162:3) {$\epsilon_{v}(e_5)$}; 
			
			\draw (e1) -- (e3r);
			\draw (e1) -- (e3l);
			\draw (e2) -- (e3l);
			\draw (e2) -- (e3m);
			\draw (e3m) -- (e3r);
			\draw (e3m) -- (e3l);
			\draw (e2) -- (e3r);
			
			\draw (e1) -- node[above right, label_edge] {$\eta\epsilon_{v}(f_{12})$} (e2);
			\draw (e2) -- node[right, label_edge] {$\eta\epsilon_{v}(f_{23})$} (e3r);
			\draw (e3r) -- node[below, label_edge] {$\eta\epsilon_{v}(f_{34})$} (e3l);
			\draw (e3l) -- node[left, label_edge] {$\eta\epsilon_{v}(f_{45})$} (e3m);
			\draw (e3m) -- node[above left, label_edge] {$\eta\epsilon_{v}(f_{15})$} (e1);
			
		\end{tikzpicture}
		
		\caption{Causal graph of a 5-valent vertex. To keep the picture clean, we  we explicitly label only the external links.}
		\label{fig:grafo_causal}
	\end{figure}
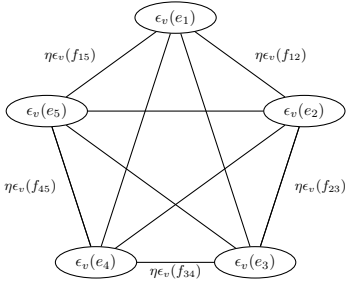
	
	For arbitrary $N_{v}^{e}$ we can construct the incidence matrix $\mathcal{A}_{v}$ associated with the graph $\mathcal{G}_{v}$ \cite{Diestel:2017, Biggs:1993}. We will call it the \textit{causality matrix of $v$}. The rows of $\mathcal{A}_{v}$ correspond to the edges of $\mathcal{G}_v$, and the columns correspond to the nodes. The matrix indicates which nodes are connected by which edges: a $1$ in an entry means that the node lies on the boundary of the corresponding edge, otherwise it is $0$. For example, the causality matrix associated with the causal graph of the five-valent vertex of Figure \ref{fig:grafo_causal} is
	
	\begin{equation}
		\begin{pmatrix}
			1 & 1 & 0 & 0 & 0 \\
			1 & 0 & 1 & 0 & 0\\
			1 & 0 & 0 & 1 & 0\\
			1 & 0 & 0 & 0 & 1\\
			0 & 1 & 1 & 0 & 0\\
			0 & 1 & 0 & 1 & 0\\
			0 & 1 & 0 & 0 & 1\\
			0 & 0 & 1 & 1 & 0\\
			0 & 0 & 1 & 0 & 1\\
			0 & 0 & 0 & 1 & 1
		\end{pmatrix}
	\end{equation}
	
	Let us write $\epsilon_{v}(e_i):=(-1)^{x_{i}}$ and $\eta\epsilon_{v}(f_{ij}):=(-1)^{k_{ij}}$, with $x_{i}, k_{ij}=0,1$. Then, taking logarithms, \eqref{wedgeor} takes the form
	
	\begin{equation}\label{newsys}
		x_{i}+x_{j}=k_{ij}
	\end{equation}
	This defines a linear system of $N_{v}^{f}$ equations with $N_{v}^{e}$ unknowns. The causality matrix is the matrix of coefficients associated with the system \eqref{newsys}.
	
	The analysis of the causal structure on the 2-skeleton and its invertibility to the 1-skeleton, at the level of a single vertex, can therefore be reduced to the study of the causal graph $\mathcal{G}_{v}$ and its incidence matrix $\mathcal{A}_{v}$ using graph theory and linear algebra over the Galois field $\mathbb{F}_{2}$ \cite{Diestel:2017, Biggs:1993, Godsil:2001, Cvetkovic:1980}.
	
	Given a single vertex $v$, the rank of the incidence matrix is $N_{v}^{e}-1$, so \eqref{newsys} has $N_{v}^{e}-1$ independent equations. The number of unknowns not fixed by the system \eqref{newsys} is $N_{v}^{e}-\text{rank}(\mathcal{A}_{v})=1$, and the system is solvable if and only if condition \eqref{cycle} is satisfied. For a solvable system we must add an extra condition to fully determine all the $x_{i}$ (correspondingly all the $\epsilon_{v}(e)$).
	
	\begin{itemize}
		
		\item If $N_{v}^{e}$ is odd, we define  
		
		\begin{equation}\label{volume}
			\delta:=\prod_{k=1}^{N_{v}^{e}}\epsilon_{v}(e_{k}).
		\end{equation}
		
		Then the system formed by \eqref{wedgeor} (equivalently \eqref{newsys}) together with \eqref{volume} is invertible, and it can show that
		
		\begin{equation}\label{solution}
			\epsilon_{v}(e_i)=\delta\prod_{k\neq i}\epsilon_{v}(f_{ik}).
		\end{equation}
		
		\item If $N_{v}^{e}$ is even, the knowledge of \eqref{volume} is not sufficient to fully solve \eqref{wedgeor}. If we change the sign of every $\epsilon_{v}(e)$, then $\delta$ remains the same. In this case, $\delta$ is a constant determined by the $\epsilon_{v}(f)$ and therefore does not provide additional information to solve the system. In order to determine a solution for the $\epsilon_{v}(e)$, a sufficient condition is the knowledge of the value of one $\epsilon_{v}(e)$. If, for example, we can determine $\epsilon_{v}(e_{1})$, then
		
		\[
		\epsilon_{v}(e_{i})=\eta\,\epsilon_{v}(e_{1})\epsilon_{v}(f_{1i}).
		\]
		
		From this we can determine the values of the remaining $\epsilon_{v}(e)$.
		
	\end{itemize}
	\subsubsection{Example: The graph $K_{6}$ with three lines removed}
	
\begin{figure}[htbp]
		\centering
		\begin{tikzpicture}[
			scale=0.5,
			transform shape,
			every node/.style={draw, ellipse, minimum width=1.5cm, inner sep=2pt},
			label_edge/.style={draw=none, rectangle, font=\small, inner sep=1pt}
			]
			
			\node (n1) at (120:4) {$\epsilon_\nu(e_1)$};
			\node (n2) at (60:4)  {$\epsilon_\nu(e_2)$};
			\node (n3) at (0:4)   {$\epsilon_\nu(e_3)$};
			\node (n4) at (-60:4) {$\epsilon_\nu(e_4)$};
			\node (n5) at (240:4) {$\epsilon_\nu(e_5)$};
			\node (n6) at (180:4) {$\epsilon_\nu(e_6)$};
			
			\draw (n1) -- (n3);
			\draw (n1) -- (n5);
			\draw (n2) -- (n4);
			\draw (n2) -- (n6);
			\draw (n3) -- (n5);
			\draw (n4) -- (n6);
			
			\draw (n1) -- node[above, label_edge]       {$\eta\epsilon_\nu(f_{12})$} (n2);
			\draw (n2) -- node[above right, label_edge] {$\eta\epsilon_\nu(f_{23})$} (n3);
			\draw (n3) -- node[below right, label_edge] {$\eta\epsilon_\nu(f_{34})$} (n4);
			\draw (n4) -- node[below, label_edge]       {$\eta\epsilon_\nu(f_{45})$} (n5);
			\draw (n5) -- node[below left, label_edge]  {$\eta\epsilon_\nu(f_{56})$} (n6);
			\draw (n6) -- node[above left, label_edge]  {$\eta\epsilon_\nu(f_{16})$} (n1);
			
		\end{tikzpicture}
		\caption{Causal graph of a 6-valent vertex. As in Figure 1, we explicitly display only the labels of the external links.}
		\label{fig:grafo_causal_5}
	\end{figure}
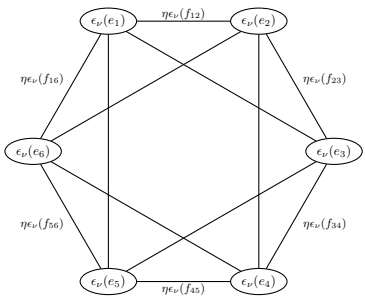

	\indent As an example of the use of the causal graph associated with a vertex $v$ and its incidence matrix, let us consider a 6-valent vertex with 12 wedges, whose boundary graph $\gamma_v$ is the $K_6$ graph with three lines removed. This graph plays an important role in some calculations of black-to-white hole tunneling \cite{DAmbrosio:2021, Christodoulou:2016}, and it provides a useful example for studying the semiclassical analysis of the EPRL-KKL spinfoam amplitude \cite{Dona:2020}. The corresponding causal graph associated with this vertex is shown in Figure \ref{fig:grafo_causal_5}. In graph theory, this graph is known as an \textit{octahedral graph}\footnote{Notice that, given a vertex $v$, the causal graph $\mathcal{G}_{v}$ and the boundary graph $\gamma_{v}$ are always topologically identical.}.
	
	\indent In this example, the causal graph $\mathcal{G}_{v}$ is a single strongly connected block (in fact, it is 4-connected). For the corresponding system of 12 equations to admit a solution, it must satisfy several consistency conditions of the form \eqref{cycle}. If condition \eqref{cycle} is satisfied for all closed cycles in $\mathcal{G}_{v}$, then the system admits two global solutions related by an overall sign change. Therefore, the knowledge of $\delta_{v}$ does not provide additional information. In order to completely determine the system, we must fix the value of one of the unknowns manually (for example, by setting $\epsilon_{v}(e_1)=+1$).
	
	\indent Because the graph has three lines removed, three equations relating certain pairs of unknowns are absent (for instance, there is no direct equation relating $\epsilon_{v}(e_1)$ and $\epsilon_{v}(e_4)$). However, due to the high connectivity of the graph, the algebraic structure already determines this relation implicitly. In particular, it can be derived by following alternative paths in the graph.
	
	\indent For example, if in $\mathcal{G}_{v}$ we can go from $\epsilon_{v}(e_1)$ to $\epsilon_{v}(e_4)$ through an intermediate node (say $\epsilon_{v}(e_2)$), we obtain the equations
	\[
	\epsilon_{v}(e_1)\epsilon_{v}(e_2)=\eta\,\epsilon_{v}(f_{12}),
	\qquad
	\epsilon_{v}(e_2)\epsilon_{v}(e_4)=\eta\,\epsilon_{v}(f_{24}).
	\]
	
	Multiplying these two relations we obtain
	\[
	\epsilon_{v}(e_1)\epsilon_{v}(e_2)^2\epsilon_{v}(e_4)
	=
	\eta^2\,\epsilon_{v}(f_{12})\epsilon_{v}(f_{24}).
	\]
	
	Since $\epsilon_{v}(e_2)^2=1$ and $\eta^2=1$, this reduces to
	\[
	\epsilon_{v}(e_1)\epsilon_{v}(e_4)
	=
	\epsilon_{v}(f_{12})\epsilon_{v}(f_{24}).
	\]
	In this way we obtain a ``virtual'' equation. Even though the corresponding edge was removed from the graph, the algebraic structure forces an effective relation between the variables. The effective coefficient of the corresponding virtual edge can therefore be written as
	\[
	\epsilon_{v}(f_{14(\mathrm{virtual})})
	=
	\epsilon_{v}(f_{12})\epsilon_{v}(f_{24}),
	\]
	noting that the dependence on $\eta$ disappears when the path is even.
	\\\\
	\indent Now let us consider the case of multiple vertices. The possibility of fully determining a consistent causal structure for the 1-skeleton from that of the 2-skeleton is affected by the results obtained above.
	
	\begin{itemize}
		
		\item Suppose we have two vertices $v_{1}, v_{2}$ connected by an edge $e$, and that $N_{v_1}^{e}$ and $N_{v_2}^{e}$ are even. In this case it is possible to determine all but one of the variables (say $\epsilon_{v}(e_1)$). All the $\epsilon_{v}(e_i)$ can then be expressed in terms of $\epsilon_{v}(e_1)$ and the $\epsilon_{v}(f)$.
		
		\item When $v_{1}$ and $v_{2}$ are connected by an edge $e$ and at least one of the two numbers, say $N_{v_1}^{e}$, is odd, the knowledge of \eqref{volume}
		\[
		\delta_{v_1}:=\prod_{k=1}^{N_{v_1}^e}\epsilon_{v_1}(e_k)
		\]
		can be used to determine the values of $\epsilon_{v_1}(e_k)$. Using the relation $\epsilon_{v_1}(e_k)=-\epsilon_{v_2}(e_k)$, we can then uniquely determine the values of all the $\epsilon_{v_2}(e_k)$.
		
	\end{itemize}
	
	Our previous results show that if the discretization contains only vertices with even valence, the causal structure of the 1-skeleton can be obtained from that of the 2-skeleton up to an undetermined quantity $\epsilon_{v}(e)$ for some $v\in V$ and $e\in\partial v$. However, if the discretization contains at least one vertex with odd valence, it is possible to fully determine the causal structure of the 1-skeleton by fixing one $\delta_{v}$ for some vertex $v$ with odd valence. Moreover, changing $\delta_{v}$ to $-\delta_{v}$ for that vertex reverses the orientations $\epsilon_{v}(e)$ of all the edges $e$. In the remainder of this article we will assume that the discretization contains at least one vertex with odd valence.
	
	\subsection{Causal structure of the boundary graph}
	
	The boundary of $\mathcal{C}$ is a graph $\gamma$ whose links are edges $e$ that belong to only one face $f$, and whose nodes $v$ are vertices that belong to exactly one internal edge. The causal structure induced on the boundary graph $\gamma$ consists of assigning a sign to each node, depending on the direction of the internal edge attached to it. We take the sign to be positive if the internal edge is future-directed and negative if it is past-directed. As shown in \cite{Bianchi:2024}, a fixed causal structure on the boundary graph $\gamma$ does not determine a unique causal structure in the bulk.
	
	The causal structure of the boundary graph can also be defined on its links. To each link we associate the product of the signs of the nodes at its endpoints. When two bulk edges that intersect the boundary share the same vertex in the bulk and form a wedge, the sign of the wedge $\epsilon_{v}(f)$ is $\eta$ times the sign of the corresponding link on the boundary. Moreover, the product of the signs of the links around any closed cycle of the boundary graph satisfies
	
	\begin{equation}\label{cycle1}
		\prod_{l\in\text{closed cycle}}\epsilon_{l}=1
	\end{equation}
	
	An assignment of signs to the links of the boundary graph defines a consistent causal structure if it allows one to consistently assign signs to the nodes. This occurs if and only if condition \eqref{cycle1} is satisfied.
	
	\section{The EPRL-KKL Model and its Causal structure}
	
	\subsection{The EPRL-KKL Spinfoam Vertex}
	The Lorentzian Engle--Pereira--Rovelli--Livine spinfoam amplitude (EPRL model) is widely regarded as the most successful spinfoam amplitude to date \cite{Perez:2012wv, R-V:2014, Engle:2007wy, Rovelli:2011eq}. The model is defined on a discretization that is dual to a triangulation. In order to establish a closer relation with the canonical approach, it is necessary to compute transition amplitudes for more general spin networks, and not only for those arising from the dual of a triangulation. The purpose of \cite{Kaminski:2010, Kaminski:2009cc} was precisely to generalize the EPRL model to arbitrary 2-complexes $\mathcal{C}$.
	
	Following \cite{Dona:2020}, consider an arbitrary 2-complex $\mathcal{C}$ and a vertex $v$ in $\mathcal{C}$ with boundary graph $\gamma_{v}$. Then $\gamma_v$ is an oriented graph with $N_{v}^{e}$ nodes and $N_{v}^{f}$ links. As in the bulk, we label the nodes of $\gamma_v$ as $a=1,\dots, N_{v}^{e}$ and the links by pairs $(ab)$. The EPRL-KKL vertex amplitude for $\gamma_v$ (as given in \cite{Ding:2010fw} and written in the form used in \cite{Dona:2020}) is
	
	\begin{equation}\label{KKLam}
		\begin{split}
			A_{\gamma_v}(j_{ab}, m_{ab}, n_{ab}) := & \int_{SL(2,\mathbb{C})} \prod_{a=2}^{N_{v}^{e}} dh_{a} \\
			& \times \prod_{(ab)} D_{j_{ab} m_{ab} j_{ab} n_{ab}}^{\gamma j_{ab}, j_{ab}}\left(h_{a}^{-1}h_{b}\right),
		\end{split}
	\end{equation}
	where $\gamma$ is the Immirzi parameter, $h_{a}\in SL(2,\mathbb{C})$, and $D^{p,k}(h)$ are the Wigner matrices of the infinite-dimensional unitary representations of the principal series, labeled by $\rho\in\mathbb{R}$ and $k\in \mathbb{Z}/2$. Each matrix $D_{j_{ab} m_{ab} j_{ab} n_{ab}}^{\gamma j_{ab}, j_{ab}}(h_{a}^{-1}h_{b})$ is associated with a wedge $(v,f)$, where $e_a$ and $e_b$ are the two edges associated with that wedge.
	
	The integrals defining the amplitude are taken over the non-compact group $SL(2,\mathbb{C})$, which makes the convergence of \eqref{KKLam} subtle. As mentioned in \cite{Dona:2020}, a sufficient condition for the amplitude to be finite is to remove one of the integrals, as done in \eqref{KKLam}, and to restrict all vertex boundary graphs $\gamma_{v}$ to be 3-link-connected. This fact justifies the restriction imposed on $\mathcal{C}$ in the previous section.
	
	The asymptotic analysis of the EPRL-KKL vertex amplitude is performed using linear combinations of \eqref{KKLam} and $SU(2)$ coherent states.\cite{Dona:2020}.
	
	The coherent EPRL-KKL spinfoam vertex amplitude is
	
	\begin{equation}\label{coherent}
		\begin{split}
			A_{\gamma_v}(j_{ab}, \vec{n}_{ab}) := & \int_{SL(2,\mathbb{C})} \prod_{a=2}^{N_{v}^{e}} dh_{a} \\
			& \times \prod_{(ab)} D_{j_{ab}, -\vec{n}_{ab}, j_{ab}, \vec{n}_{ba}}^{\gamma j_{ab}, j_{ab}}\left(h_{a}^{-1}h_{b}\right),
		\end{split}
	\end{equation}
	
	where
	
	\begin{equation}\label{wigcoh}
		\begin{split}
			D_{j_{ab} -\vec{n}_{ab} j_{ab} \vec{n}_{ba}}^{\gamma j_{ab}, j_{ab}}(h_{a}^{-1}h_{b}) = & \exp(-2ij_{ab}\Phi_{ab})\dfrac{d_{j_{ab}}}{\pi} \\
			& \times \int_{\mathbb{C}P^{1}} d\mu(z_{ab}^{A}) 
			\dfrac{\exp(S_{ab})}{\|h_{a}^{T}z_{ab}\|^{2}\|h_{b}^{T}z_{ab}\|^2},
		\end{split}
	\end{equation}
	
	and
	
	\begin{equation}\label{action}
		S_{ab}:=j_{ab}\ln\dfrac{\langle h_{a}^{T}z_{ab}|\bar{\zeta}_{ab}\rangle^{2}\left[\bar{\zeta}_{ba}|h_{b}^{T}z_{ab}\rangle^{2}\right.}{\|h_{a}^{T}z_{ab}\|^{2}\|h_{b}^{T}z_{ab}\|^{2}}+i\gamma j_{ab}\ln\dfrac{\|h_{b}^{T}z_{ab}\|^{2}}{\|h_{a}^{T}z_{ab}\|^{2}}.
	\end{equation}
	
	The matrices $D^{\rho, k}$ are formulated using minimal spin eigenvalues, together with the unit vectors $\vec{n}_{ab}$ associated to the links $(ab)$ of $\gamma_v$ and the auxiliary spinors $z_{ab}$. The set of vectors provides the fundamental building blocks for constructing coherent intertwiners, which characterize different types of boundary geometries. The spinors $|z_{ab}\rangle$ give the homogeneous realization of the infinite irreducible representations of $SL(2,\mathbb{C})$. The spinorial map is defined by
	$$\left.|z\right]:=\epsilon|\bar{z}\rangle,\quad \epsilon = i\sigma_2 = 
	\begin{pmatrix} 
		0 & 1 \\ 
		-1 & 0 
	\end{pmatrix},$$
	and the integral is performed over $\mathbb{C}P^{1}$ with measure $d\mu(z)=(i/2)\left[z|dz\rangle\right.\wedge \left[\bar{z}|d\bar{z}\rangle\right.$. Our approach follows the conventions and expressions established in \cite{Dona:2020, Dona:2019}.
	
	The complete asymptotic analysis of \eqref{coherent} was carried out in \cite{Dona:2020}. The asymptotic limit is taken as the limit of large quantum numbers $j$. This is usually studied by making the transformation $j_{ab}\rightarrow\lambda j_{ab},\quad\forall (ab)$ in the regime $\lambda\rightarrow\infty$. In this limit, the partition function of $\mathcal{C}$ takes the general form:
	
	$$\int g(z)e^{\lambda f(z)}d^{n}z.$$
	This allows the use of the stationary phase method to found an approximation of the partition function in the limit of large $\lambda$. The procedure implemented in \cite{Dona:2020} allows to solve the critical point equations using elementary trigonometry and to determine the explicit form of the group elements $h_{a}$ at the critical points through an appropriate gauge choice. Further details can be found in \cite{Dona:2020}.
	
	For non-degenerate Lorentzian boundary conditions, the results show that the asymptotic behavior of the amplitude depends on the structure and connectivity of the boundary graph $\gamma_{v}$\footnote{An $n-$modular graph is defined as a graph that can be divided in $n$ subgraphs, such that every node in each subgraph is connected to at most one node in another subgraph}.
	
	\begin{itemize}
		
		\item For 1-modular graphs and irreducible amplitudes, there are two different critical points related by a parity transformation. In the asymptotic limit, the vertex amplitude takes the form
		
		\begin{equation}\label{asin1}
			A_{\gamma_v}\sim\dfrac{\Omega^{+}e^{i\lambda\gamma S_{\gamma_{v}}}}{\sqrt{\text{det}\left(-H^{+}\right)}}+\dfrac{\Omega^{-}e^{-i\lambda\gamma S_{\gamma_{v}}}}{\sqrt{\text{det}\left(-H^{-}\right)}} ,
		\end{equation}
		
		with $S_{\gamma_v}:=\sum_{ab}j_{ab}\theta_{ab}$ a generalization of the Regge action for an arbitrary cellular decomposition \cite{Dona:2020}. Here $H^{\pm}$ is the value of the Hessian of the total action $S(h,z):=\sum_{(ab)}S_{ab}$ at the two critical points, $\Omega^{\pm}$ denotes the value of the spinorial measure at the two critical points \cite{Dona:2020, Barrett:2011xa, Barrett:2010}, and $\theta_{ab}$ is the dihedral angle between the boundary polyhedra in $a$ and $b$.
		
		\item For $n$-modular graphs ($n>1$), or if the amplitude is reducible using recoupling theory, there are $2^{n}$ pairs of critical points, one pair for each module (or subgraph). In this case
		
		\begin{equation}\label{asinn}
			A_{\gamma_v}\sim\sum_{\epsilon=\pm}\sum_{\eta=\pm}\dfrac{\Omega^{(\epsilon, \eta_{i})}}{\sqrt{\text{det}\left(-H^{\epsilon, \eta_{i}}\right)}}e^{i\epsilon\lambda\gamma S_{\gamma_v}^{\eta_{i}}},
		\end{equation}
		
		where
		
		\begin{equation}
			S_{\eta_i\gamma_v} := \sum_{(ab)\notin\cup I_i} j_{ab}\hat{\theta}_{ab} + \sum_{i=1}^{n-1} \sum_{(ab)\in I_i} j_{ab}\hat{\theta}_{ab}(\eta_i).
		\end{equation}
		
		The variable $\eta_i$ denotes an additional degree of freedom associated with the links that connect different modules, and $I_{i}$ is the set of links connecting different modules ($i=1,\dots, n-1$).
		
	\end{itemize}
	
	For the EPRL-KKL amplitude, there is no evidence that $H^{+}=\overline{H^{-}}$. Moreover, there is currently no analytical proof that the Hessian associated with a general boundary graph is non-degenerate \cite{Dona:2020}.
	
	\subsection{Causal structure in the EPRL-KKL model}
	The EPRL-KKL model can be obtained as a quantization of a constrained BF theory satisfying the linear simplicity constraint, discretized on an arbitrary oriented 2-complex $\mathcal{C}$. This implies, among other things, the assignment of an element $B^{IJ}$ of the Lie algebra $sl(2,\mathbb{C})$ to each face $f\in F$ of $\mathcal{C}$. Details of the construction can be found in \cite{Ding:2010fw}. For the purposes of this article, we are interested in the analysis of the linear simplicity constraint:
	
	\begin{itemize}
		\item There exists a unit vector $(n_{e})_I$ for each $e\in E$ such that, for all $f\in\partial e$,
		\begin{equation}\label{linear}
			(n_{e})_{I}B_{f}^{IJ}=0 .
		\end{equation}
	\end{itemize}
	
	The linear simplicity constraint establishes two conditions: the existence of a unit vector $n_{e}$ for each edge $e$, and the relation $(n_{e})_{I}B_{f}^{IJ}=0$. The vectors $n_{e}$ allow to associate a causal structure to $\mathcal{C}$. Let us consider an edge $e$ and two vertices $v_{1}, v_{2}$ such that $(e, v_1)$ and $(e, v_2)$ are admissible pairs. Then the function
	
	\[
	\mathcal{X}(e, v_1):=n_{e}, \qquad \mathcal{X}(e, v_2):=-n_{e}
	\]
	
	defines a causal structure on $\mathcal{C}$. There is, however, an ambiguity in this choice, since nothing prevents us from instead defining
	
	\[
	\mathcal{Y}(e, v_1):=-n_{e}, \qquad \mathcal{Y}(e, v_2):=n_{e}.
	\]
	
	According to Section 2, these correspond to two inequivalent causal structures which are nevertheless related by a time-reversal operation. Moreover, the linear simplicity constraint is satisfied for both assignments $n_{e}$ and $-n_{e}$ for each edge. This implies that both histories contribute to the spinfoam path integral in the EPRL-KKL model.
	
	Given a single vertex $v$, the orientations of the wedges associated with $v$ may satisfy condition \eqref{cycle} for $\eta=+1$, $\eta=-1$, or may fail to satisfy the condition altogether. Then, as in the EPRL model \cite{Bianchi:2024}, the EPRL-KKL amplitude can be written as a sum of three contributions
	
	\begin{equation}\label{terms}
		A_{\gamma_v}=\sum_{\substack{[\epsilon_{ab}]\\\eta=+1}}A_{v}^{\epsilon_{ab}}+\sum_{\substack{[\epsilon_{ab}]\\\eta=-1}}A_{v}^{\epsilon_{ab}}+\sum_{\substack{[\epsilon_{ab}]\\\text{non-causal}}}A_{v}^{\epsilon_{ab}}.
	\end{equation}
	The first two terms on the right-hand side correspond to configurations of wedges satisfying \eqref{cycle} for $\eta=+1$ and $\eta=-1$, respectively. The last term corresponds to configurations that do not satisfy \eqref{cycle} and therefore do not define a causal structure for $\mathcal{C}$, as introduced in Section 2.

	\section{The Causal EPRL-KKL Spinfoam Vertex}
	
	Motivated by the analysis of the previous sections, we are led to consider an orientation-dependent extension of the EPRL-KKL spinfoam model, defining a causal spinfoam amplitude. Such amplitudes describe causal transition processes and admit an interpretation as quantum gravity analogues of the Feynman propagator in quantum field theory. This construction has already been proposed for the EPRL model in \cite{Bianchi:2024, Bianchi:2026rjd}.
	
	According to the analysis in \cite{Dona:2020}, for non-degenerate Lorentzian boundary conditions describing a Regge geometry, the quantity
	
	$$
	\ln\dfrac{\|h_{b}^{T}z_{ab}\|^{2}}{\|h_{a}^{T}z_{ab}\|^{2}}
	$$
	is, essentially, the 4D dihedral angle associated with a wedge $(v,f)$, as determined by the 3D Regge geometry on the boundary of the vertex. The analysis of the previous section suggests that the causal structure of the 2-skeleton of the EPRL-KKL is encoded in this quantity. We can follow a brute force procedure and defining a causal vertex amplitude in a way similar to \cite{Bianchi:2024}:
	
	\begin{equation}\label{causalvertex}
		\begin{aligned}
			A_{\gamma_v}^{\epsilon}(j_{ab}, \vec{n}_{ab}) := & \prod_{(ab)} e^{-2ij_{ab}\Phi_{ab}} \dfrac{d_{j_{ab}}}{\pi} \int_{SL(2,\mathbb{C})} \prod_{a=2}^{N_{v}^{e}} dh_{a} \\
			& \times \int_{\mathbb{C}P^{1}} d\mu(z_{ab}^{A}) \dfrac{\exp(S_{ab})}{\|h_{a}^{T}z_{ab}\|^{2}\|h_{b}^{T}z_{ab}\|^2} \\
			& \times \Theta\left[\epsilon_{ab}\gamma j_{ab}\ln\dfrac{\|h_{b}^{T}z_{ab}\|^{2}}{\|h_{a}^{T}z_{ab}\|^{2}}\right],
		\end{aligned}
	\end{equation}
	where $\Theta$ denotes the step function. The EPRL-KKL amplitude is recovered by summing over all possible assignments of orientations to the wedges,
	
	\begin{equation}\label{sum}
		A_{v}(j_{ab}, \vec{n}_{ab}) = \sum_{[\epsilon_{ab}]}A_{\gamma_v}^{\epsilon}(j_{ab}, \vec{n}_{ab}) .
	\end{equation}
	The sum contains $2^{M}$ terms, where $M$ is the number of wedges in the vertex $v$.
	
	Interpreting $\epsilon_{ab}$ as the causal orientation of the wedge $(ab)$, an assignment of orientations to each wedge defines a proper causal structure when condition \eqref{cycle} is satisfied. Hence $A_{v}^{\epsilon}$ can be called a causal amplitude whenever the set $\{\epsilon_{ab}\}$ satisfies \eqref{cycle} for every cycle in $v$.
	
	Equation \eqref{causalvertex} is an \textit{ad hoc} proposal formulated using the coherent states basis and a step function put by hand. Following \cite{Bianchi:2026rjd}, we can provide an exact realization of \eqref{causalvertex} by replacing in \eqref{KKLam} the Wigner matrices $D^{\gamma j, j}(h)$ with the so-called Toller matrices (introduced in \cite{Bianchi:2026rjd}):
	
	\begin{equation}\label{BCG}
		\begin{aligned}
			A_{\gamma_v}^{\epsilon}(j_{ab}, m_{ab}, n_{ab}):=&\int_{SL(2,\mathbb{C})}\prod_{a=2}^{N_{v}^{e}}dh_{a}\\
			&\times \prod_{(ab)}T_{j_{ab} m_{ab} j_{ab} n_{ab}}^{\epsilon_{ab}, \gamma j_{ab}, j_{ab}}\left(h_{a}^{-1}h_{b}\right)
		\end{aligned}
	\end{equation}
	
	This defines a generalized Bianchi-Chen-Gamonal (BCG) amplitude associated with an arbitrary 2-complex $\mathcal{C}$.
	
	Using the results of \cite{Dona:2020}, we can determine the asymptotic limit of \eqref{BCG}. The causal amplitude of a coherent state peaked on a non-degenerate Lorentzian Regge geometry is
	
	\begin{equation}\label{amcoh}
		\langle A_{v}^{\epsilon}|j_{ab},\zeta_{ab}\rangle=\int_{SL(2,\mathbb{C})}\prod_{a=2}^{N_{v}^{e}}dh_{a}\prod_{(ab)}T_{j_{ab} \zeta_{ab} j_{ab} J\zeta_{ab}}^{\epsilon_{ab}, \gamma j_{ab}, j_{ab}}\left(h_{a}^{-1}h_{b}\right).
	\end{equation}
	where $T_{j_{ab} \zeta_{ab} j_{ab} J\zeta_{ab}}^{\epsilon_{ab}, \gamma j_{ab}, j_{ab}}\left(h_{a}^{-1}h_{b}\right)$ is the Toller matrix in the coherent spinorial basis \cite{Bianchi:2026rjd}. Using the conventions of \cite{Dona:2020}, \cite{Dona:2019} and equation (3) of \cite{Bianchi:2026rjd}, the Toller matrix in \eqref{amcoh} reads
	
	\begin{equation}\label{tollercoh}
		\begin{aligned}
			T_{j_{ab} \zeta_{ab} j_{ab} J\zeta_{ab}}^{\epsilon_{ab}, \gamma j_{ab}, j_{ab}}\left(h_{a}^{-1}h_{b}\right)= & \dfrac{d_{j_{ab}}}{\pi}\int_{\mathbb{C}P^{1}} \dfrac{d\mu(z_{ab}^{A})}{\|h_{a}^{T}z_{ab}\|^{2}\|h_{b}^{T}z_{ab}\|^2}\\
			& \times G_{\epsilon_{ab}, \gamma j_{ab}, j_{ab}}\left[B_{ab}(h,z)\right]\\
			& \times e^{j_{ab}\psi_{ab}(h,z,\zeta)}e^{i\rho_{ab}B_{ab}(h,z)}
		\end{aligned}
	\end{equation}
	where
	
	\begin{equation}\label{G}
		G_{\epsilon_{ab}, \gamma j_{ab}, j_{ab}}\left[x\right]=\Theta(\epsilon_{ab}x)+\epsilon_{ab}\delta^{\gamma j_{ab}, j_{ab}}(x),
	\end{equation}
	with
	
	\begin{equation}\label{B}
		B_{ab}(h,z):=\ln\left[\dfrac{\|h_{b}^{T}z_{ab}\|^2}{\|h_{a}^{T}z_{ab}\|^2}\right],
	\end{equation}
	and
	
	\begin{equation}\label{psi}
		\psi_{ab}(h, z, \zeta):=\ln\dfrac{\langle h_{a}^{T}z_{ab}|\bar{\zeta}_{ab}\rangle^{2}\left[\bar{\zeta}_{ba}|h_{b}^{T}z_{ab}\rangle^{2}\right.}{\|h_{a}^{T}z_{ab}\|^{2}\|h_{b}^{T}z_{ab}\|^{2}}.
	\end{equation}
	The Dirac delta distribution $\delta^{(\rho, j)}(x)$ is defined as\cite{Bianchi:2026rjd}:
	
	\begin{equation}\label{dirac}
		\delta^{(\rho, j)}(x):=\sum_{n=0}^{2j}\dfrac{c_{n+1}^{(\rho, j)}}{(n+1)!}\dfrac{d^{n}}{dx^{n}}\delta(x).
	\end{equation}
	The coefficients $c_{n+1}^{(\rho, j)}$	are given by:
	
	\begin{equation}\label{coeff}
		c_{n+1}^{(\rho, j)}:=\dfrac{d^{n}}{d\tilde{s}^{n}}\left(\dfrac{\Gamma(-i-is)\Gamma(i-i\tilde{s}+1)}{\Gamma(-j-i\tilde{s})\Gamma(j-is+1)}\right),
	\end{equation}	
	where $\Gamma(x)$ is the Gamma function.
	
	Notice that we essentially recover expression \eqref{causalvertex}, except for the presence of the additional Dirac delta distribution $\delta^{(\rho, j)}(x)$ with support at $x=0$.
	
	For non-degenerate Lorentzian boundary conditions, the function $B_{ab}(h,z)$ evaluated at the critical points is \cite{Dona:2019}
	
	$$
	B_{ab}(h,z)=\pm\frac{\theta_{ab}}{2}\neq 0,
	$$
	and therefore the Dirac delta distribution vanishes. The behavior of \eqref{amcoh} depends on the structure of the boundary graph:
	
	\begin{itemize}
		
		\item For 1-modular graphs, the situation is similar to that of a 4-simplex \cite{Dona:2019}. There are two critical points, and only one of them, $+\theta_{ab}/2$, contributes to an oscillatory amplitude. The other point, $-\theta_{ab}/2$, does not contribute to the asymptotics because of the $\Theta$ function. Then the causal vertex amplitude \eqref{amcoh} takes the form
		
		\begin{equation}\label{asym1}
			\langle A_{v}^{\epsilon}|j_{ab},\zeta_{ab}\rangle\sim \dfrac{\Omega^{(+)}e^{i\lambda\gamma S_{\gamma_v}}}{\sqrt{\text{det}(-H^{+})}}
		\end{equation}
		
		\item For $n$-modular graphs, with $n\geq 2$, each module has a pair of critical points, and within each pair only one produces an oscillatory behavior. In the semiclassical regime the amplitude becomes
		
		\begin{equation}\label{asym2}
			\langle A_{v}^{\epsilon}|j_{ab},\zeta_{ab}\rangle\sim \sum_{\eta_i=\pm}\dfrac{\Omega^{(\eta_i)}e^{i\lambda\gamma S_{\gamma_v}^{\eta_i}}}{\sqrt{\text{det}(-H^{\eta_i})}}
		\end{equation}
		
	\end{itemize}
	
	\section{Discussion}
	\subsection{Changes of signature}
	
	As mentioned in section 3.2, the EPRL-KKL model can be considered a sum with three contributions:
	$$A_{\gamma_v}=\sum_{\substack{[\epsilon_{ab}]\\\eta=+1}}A_{v}^{\epsilon_{ab}}+\sum_{\substack{[\epsilon_{ab}]\\\eta=-1}}A_{v}^{\epsilon_{ab}}+\sum_{\substack{[\epsilon_{ab}]\\\text{non-causal}}}A_{v}^{\epsilon_{ab}},$$
	associated with causal configurations with $\eta=+1$, $\eta=-1$ and configurations that do not satisfy the condition \eqref{cycle}.
	
	In the same way as in \cite{Bianchi:2024}, the $2^{n}$ critical points in the asymptotic limit of an $n$-modular graph can be interpreted as arising from the terms $\sum_{\substack{[\epsilon_{ab}]\\\eta=+1}}A_{v}^{\epsilon_{ab}}$ and $\sum_{\substack{[\epsilon_{ab}]\\\eta=-1}}A_{v}^{\epsilon_{ab}}$, associated with changes in the signature $\eta$.
	
	The presence of two critical points in the EPRL model is known as the \textit{cosine problem} \cite{Engle:2023b, Engle:2013, Engle:2016}. It has been argued that the cosine problem can lead to difficulties when analyzing discretizations containing more than one vertex \cite{Engle:2023b}. These issues could potentially be exacerbated in the EPRL-KKL vertex due to the presence of multiple pairs of critical points for some $n$-modular graphs. As in the causal EPRL amplitude introduced in \cite{Bianchi:2024} and \cite{Bianchi:2026rjd}, the causal amplitude presented here provides an alternative that may help overcome these difficulties.
	
	The cosine problem, together with the interpretation of the two critical points in the asymptotic analysis of the EPRL amplitude as arising from changes in the convention $\eta$, raises an additional conceptual issue. If we interpret $e^{+iS_{\text{Regge}}}$ and $e^{-iS_{\text{Regge}}}$ as originating from two different choices of $\eta$, and if both choices are allowed and physically equivalent, then insisting on retaining only the term $e^{+iS_{\text{Regge}}}$ amounts to selecting a single signature, namely $\eta=+1$. This would appear to contradict the common understanding that the two conventions $\eta=+1$ and $\eta=-1$ are physically equivalent, and then must be included in a gravitational path integral.
	
	This analysis suggests, in our view, that the interpretation of the two contributions $e^{+iS_{\text{Regge}}}$ and $e^{-iS_{\text{Regge}}}$ as arising solely from different choices of $\eta$ requires further clarification.
	
	\subsection{Irregular light-cone structures and the causal vertex}
	
	Expression \eqref{terms} shows that the EPRL-KKL amplitude can be written as a sum of three contributions. The causal vertex amplitude defined as
	
	\begin{equation}\label{causal+}
		A_{\gamma_v}^{+}:=\sum_{\substack{[\epsilon_{ab}]\\\eta=+1}}A_{v}^{\epsilon_{ab}}
	\end{equation}
	has the asymptotic limit $\sim e^{iS_{\text{Regge}}}$.
	
	On the other hand, the Lorentzian Regge action can become complex due to the presence of complex deficit angles \cite{Sorkin:2019}. These correspond to configurations with an irregular light-cone structure and describe changes in the topology of the spatial manifold during its temporal evolution \cite{Asante:2021}. The precise role of these configurations in the EPRL or the EPRL-KKL model is not yet fully understood. The analysis of \cite{Asante:2021} suggests that these irregular light-cone structures could play a role in large-scale physics and might, in principle, need to be excluded.
	
	In a discretization containing more than one vertex, replacing the EPRL-KKL amplitude with its causal alternative $A_{v}^{+}$ at each vertex could provide a mechanism to prevent the appearance of such irregular light-cone structures. A more general proposal is to exclude all non-causal configurations and to study instead the amplitude
	
	$$
	\sum_{\substack{[\epsilon_{ab}]\\\eta=+1}}A_{v}^{\epsilon_{ab}}+\sum_{\substack{[\epsilon_{ab}]\\\eta=-1}}A_{v}^{\epsilon_{ab}} .
	$$
	
	This restriction should be understood as a proposal motivated by the semiclassical analysis given in the previous section.
	
	\section{Conclusions and open issues}
	
	In this article we analyzed the role of causality in generalized spinfoam models defined on an arbitrary 2-complex $\mathcal{C}$. We proposed a definition of causal structure for an arbitrary 2-complex $\mathcal{C}$ and studied how such a structure can induce causal orientations on the 1-skeleton and the 2-skeleton. We found that not all orientation assignments on the 2-skeleton correspond to proper causal structures. We identified an interesting criterion to determine when a given orientation on the 2-skeleton defines a consistent causal structure on the 1-skeleton. This analysis naturally involves tools from graph theory and linear algebra over the Galois field $\mathbb{F}_2$.
	
	There remain, however, several open issues that deserve further investigation:
	
	\begin{itemize}
		
		\item The EPRL-KKL model defines a vertex amplitude that can be divergent. This divergence is typically regulated by removing one of the integrals in the definition of the amplitude and by restricting the boundary graph to be 3-link connected. The finiteness of the generalized causal vertex amplitude introduced here remains an open question that should be explored.
		
		\item The asymptotic analysis of the generalized causal amplitude presented here focused on boundary data associated with a non-degenerate Lorentzian Regge geometry. However, the analysis of \cite{Dona:2019} suggests that the EPRL-KKL amplitude admits critical points corresponding to different types of boundary geometries. These include conformal twisted geometries and non-flat-embeddable 3D Regge geometries. It would therefore be important to study the asymptotic behavior of the causal amplitude for these configurations too.
		
		\item It would also be interesting to investigate whether the causal vertex amplitudes introduced here, as well as those proposed in \cite{Bianchi:2024} and \cite{Bianchi:2026rjd}, provide sufficient conditions to prevent the emergence of irregular light-cone structures when discretizations with more than one vertex are considered.
		
		\item The EPRL-KKL model has been applied in several contexts, including spinfoam cosmology, black-to-white hole tunneling, and studies of renormalization in spinfoams \cite{Christodoulou:2024, Bianchi:2018rem, DAmbrosio:2021, Soltani:2021, Bianchi:2010cosmo, Bahr:2016a, Bahr:2016b, Bahr:2018}. It would be interesting to investigate the impact of the generalized causal amplitude introduced here in these settings. In particular, one intriguing possibility is the existence of quantum tunneling processes between different causal structures.
		
	\end{itemize}

	\section*{Acknowledges}
	The author would like to thank to his PhD Supervisor José Zapata for some helpful discussions, and to Mauricio Gamonal for some conversations and clarifications about the new causal vertex introduced in \cite{Bianchi:2026rjd}.
	
	
	Finally, the author wishes to express his very special gratitude to \textbf{Maria Fernanda Clever Uribe}, for reminding him, a stellar cycle ago, that the atoms composing our gazes were forged within the same stellar nucleus. This work, as the result of that ancient and persistent observation of the universe, is dedicated to her.
	
	This research was supported by the Programa de Apoyo a Proyectos de Investigación e Innovación Tecnológica (PAPIIT), DGAPA UNAM, through projects IN114723 and IN119726.

	\appendix
	\titleformat{\section}
	{\normalfont\bfseries\large}
	{Appendix \thesection.}{0.5em}{}
	\numberwithin{equation}{section}
	

\end{document}